\documentclass[aps, prb, reprint, superscriptaddress]{revtex4-1}
\usepackage[english]{babel}
\usepackage{amsmath,amsthm}
\usepackage{amsfonts}
\usepackage{xspace}
\usepackage{bm}
\usepackage{booktabs}
\usepackage{graphicx}
\usepackage{dcolumn}% Align table columns on decimal point
\usepackage[mathlines]{lineno}% Enable numbering of text and display math
\usepackage[colorlinks]{hyperref}  
\usepackage{gensymb}
\usepackage{epstopdf}
\usepackage{newfloat}
\usepackage{color,soul}
\DeclareFloatingEnvironment[name={Figure S}]{suppfigure}
\DeclareFloatingEnvironment[name={S}]{suppEquation}
\emergencystretch=\maxdimen
\begin{document}

\title{Damping enhancement in coherent ferrite/insulating-paramagnet bilayers}%

\author{Jacob J. Wisser}%
\affiliation{ 
Department of Applied Physics, Stanford University, Stanford, CA, USA
}%
\author{Alexander J. Grutter}%
\affiliation{ 
NIST Center for Neutron Research, Gaithersburg, MD, USA
}%
\author{Dustin A. Gilbert}%
\affiliation{ 
Department of Materials Science and Engineering, University of Tennessee, Knoxville, TN, USA
}%
\author{Alpha T. N'Diaye}%
\affiliation{ 
Advanced Light Source, Lawrence Berkeley National Laboratory, Berkeley, CA, USA
}%
\author{Christoph Klewe}%
\affiliation{ 
Advanced Light Source, Lawrence Berkeley National Laboratory, Berkeley, CA, USA
}%
\author{Padraic Shafer}%
\affiliation{ 
Advanced Light Source, Lawrence Berkeley National Laboratory, Berkeley, CA, USA
}%
\author{Elke Arenholz}%
\affiliation{ 
Advanced Light Source, Lawrence Berkeley National Laboratory, Berkeley, CA, USA
}%
\affiliation{
Cornell High Energy Synchrotron Source, Ithaca, NY, USA
}
\author{Yuri Suzuki}%
\affiliation{ 
Department of Applied Physics, Stanford University, Stanford, CA, USA
}%
\author{Satoru Emori}%
\email{
semori@vt.edu
}
\affiliation{ 
Department of Physics, Virginia Tech, Blacksburg, VA 24061, USA%\\This line break forced with \textbackslash\textbackslash
}%
\date{\today}% It is always \today, today,

% ----------------------------------------------------------------
\begin{abstract}
High-quality epitaxial ferrites, such as low-damping MgAl-ferrite (MAFO), are promising nanoscale building blocks for all-oxide heterostructures driven by pure spin current. However, the impact of oxide interfaces on spin dynamics in such heterostructures remains an open question. 
Here, we investigate the spin dynamics and chemical and magnetic depth profiles of 15-nm-thick MAFO coherently interfaced with an isostructural $\approx$1-8-nm-thick overlayer of paramagnetic CoCr$_2$O$_4$ (CCO) as an all-oxide model system.  Compared to MAFO without an overlayer, effective Gilbert damping in MAFO/CCO is enhanced by a factor of $>$3, irrespective of the CCO overlayer thickness. We attribute this damping enhancement to spin scattering at the $\sim$1-nm-thick chemically disordered layer at the MAFO/CCO interface, rather than spin pumping or proximity-induced magnetism. 
Our results indicate that damping in ferrite-based heterostructures is strongly influenced by interfacial chemical disorder, even if the thickness of the disordered layer is a small fraction of the ferrite thickness. 
\end{abstract}
\maketitle
% ----------------------------------------------------------------

\section{Introduction}\label{sec:intro}
Emerging spintronic device schemes leverage magnon spin currents in electrically insulating magnetic oxides (e.g., ferrites), unaccompanied by dissipative motion of electrons, for computing and communications applications~\cite{Hoffmann2015a, Chumak2015}. Low-dissipation spintronic devices become particularly attractive if insulating ferrite thin films with low magnetic damping can serve as sources of magnon spin currents. Such low-damping ferrites include not only epitaxial garnet ferrites (e.g., YIG)~\cite{DAllivyKelly2013,Du2015,Chang2014,Onbasli2014,Lustikova2014,Howe2015,Tang2016, Hauser2016, Talalaevskij2017} that have been widely used in studies of insulating spintronics~\cite{Chumak2015,DAllivyKelly2013,Du2015,Heinrich2011,Jungfleisch2015,Zhou2016a,Holanda2017}, but also coherently strained epitaxial spinel ferrites~\cite{Emori2017,Singh2017, Emori2018a} with crucial technical advantages over garnets, such as lower thermal budget for crystallization, higher magnon resonance frequencies, and potential to be integrated coherently with other spinels and perovskites with various functionalities~\cite{Ramesh2007,Zubko2011,hwang2012,Varignon2018}. 

In general, low-damping ferrite thin films must be interfaced with other materials to realize spintronic devices. It is therefore essential to understand whether and how damping in the ferrite is impacted by the proximity to another material. For instance, to convert between electronic and magnonic signals through direct and inverse spin Hall or Rashba-Edelstein effects~\cite{Sinova2015}, the low-damping ferrite needs to be interfaced with a nonmagnetic metal with strong spin-orbit coupling. Spin transport and enhanced damping through spin pumping~\cite{Tserkovnyak2002} in ferrite/spin-orbit-metal structures has already been extensively studied~\cite{DAllivyKelly2013,Du2015,Heinrich2011,Jungfleisch2015,Zhou2016a,Holanda2017,Emori2018}. Moreover, the low-damping ferrite can be interfaced with an insulating antiferromagnetic or paramagnetic oxide, in which signals can be transmitted as a pure magnon spin current~\cite{Wang2014g,Hahn2014a,Shiomi2014,Moriyama2015a,Takei2015,Qiu2016,Prakash2016,Lin2016a,Khymyn2016,Rezende2016,Okamoto2016, Wu2018, Cramer2018, Guo2018,Qiu2018}. While interfacing low-damping ferrites with insulating anti/paramagnetic oxides has enabled prototypes of magnon spin valves~\cite{Wu2018, Cramer2018, Guo2018}, the fundamental impact of insulating oxide interfaces on spin dynamics has remained mostly unexplored. In particular, it is an open question whether or how damping of the ferrite is enhanced from spin dissipation within the bulk of the adjacent anti/paramagnetic oxide or from spin scattering at the oxide interface. 

Here, we investigate how room-temperature magnetic damping in epitaxial ferrimagnetic spinel MgAl-ferrite (MgAl$_{1/2}$Fe$_{3/2}$O$_4$, MAFO) is impacted when interfaced with an overlayer of insulating paramagnetic spinel  CoCr$_2$O$_4$ (CCO)~\cite{Heuver2015, Menyuk1962}. This epitaxial MAFO/CCO bilayer is an isostructural model system, possessing a coherent interface with continuous crystal lattices between the spinel ferrite and paramagnet. We find that the presence of MAFO/CCO interface increases damping by more than a factor of $>$3 compared to MAFO without an overlayer. We attribute this damping enhancement -- which is comparable to or greater than spin pumping effects reported for ferrite/spin-orbit-metal bilayers -- to spin scattering by the ultrathin ($\sim$1 nm) chemically disordered layer at the MAFO/CCO interface. Our findings show that spin scattering at oxide interfaces has a profound influence on damping, even when the chemically disordered layer is a small fraction of the total magnetic layer thickness. 

%%%
\begin{figure*}[tb]
  \includegraphics [width=2.00\columnwidth] {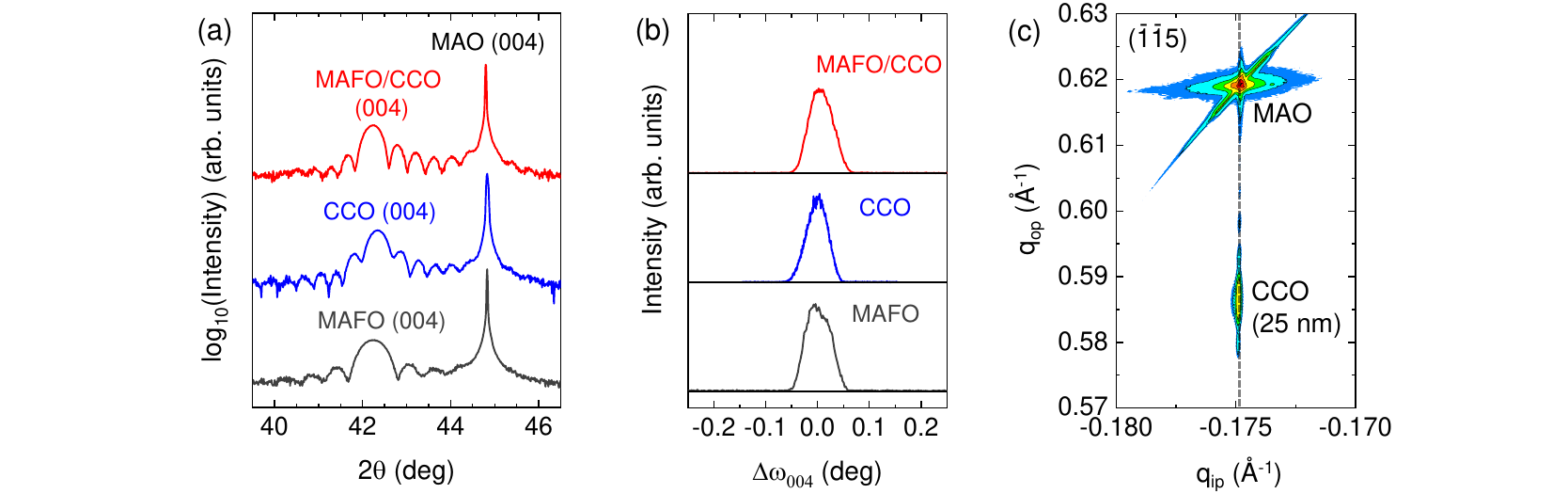}
  \centering
  \caption{\label{fig:XRD}
    (a) 2$\theta$-$\omega$ scans of epitaxial MAFO(15 nm), CCO(25 nm), and MAFO(15 nm)/CCO(8 nm). The data are offset for clarity. (b) Rocking curve scans about the (004) film peak for the films shown in (a). (c) Reciprocal space map of epitaxial CCO(25 nm) coherently strained to the MAO substrate.} 
\end{figure*}
%%%

\section{Film Growth and Structural Characterization}\label{sec:film}
Epitaxial thin films of 15-nm-thick MAFO interfaced with 1.3-8 nm of CCO overlayer were grown on as-received single-crystal MgAl$_2$O$_4$ (MAO) substrates via pulsed laser deposition. A KrF 248 nm laser was incident on stoichiometric targets of MAFO and CCO with fluences of $\approx$ 1.5 J/cm$^2$ and $\approx$ 1.3 J/cm$^2$, respectively. Both films were grown in 10 mTorr (1.3 Pa) O$_2$ and were cooled in 100 Torr (13 kPa) O$_2$. MAFO films were grown at 450 $\degree$C, whereas CCO films were deposited at 300 $\degree$C in an attempt to minimize intermixing between the MAFO and CCO layers. 
These growth temperatures, much lower than $>$700 $\degree$C typically required for epitaxial garnets~\cite{DAllivyKelly2013,Du2015,Chang2014,Onbasli2014,Lustikova2014,Howe2015,Tang2016, Hauser2016, Talalaevskij2017}, are sufficient to fully crystallize MAFO and CCO. The low crystallization temperatures of the spinels offer an advantage over the oft-studied garnets, with more opportunities for isostructural integration with coherent interfaces. 
The MAFO films exhibit a room-temperature saturation magnetization of $\approx$100 kA/m and a Curie temperature of $\approx$400 K~\cite{Emori2018a}.
To obtain consistent ferromagnetic resonance results, MAFO films were grown and subsequently characterized by ferromagnetic resonance (FMR) ex-situ; after surface cleaning with ultrasonication in isopropanol, CCO overlayers were then deposited as described above. Growth rates were calibrated via X-ray reflectivity.

Our structural characterization of MAFO and CCO shows high-quality, coherently strained films. In symmetric 2$\theta$-$\omega$ X-ray diffraction scans, only peaks corresponding to the $(00\ell)$ reflections are observed, indicating that the films are highly epitaxial. Additionally, as seen in Fig.~\ref{fig:XRD}(a), Laue oscillations around the (004) Bragg reflections in both single-layer MAFO and CCO layers as well as MAFO/CCO bilayers denote smooth interfaces. Furthermore, MAFO, CCO, and MAFO/CCO samples all exhibit essentially the same film-peak rocking curve widths (FWHM) of $\approx$0.06$^\circ$ (Fig.~\ref{fig:XRD}(b)). 
Reciprocal space mapping of the $(\Bar{1}\Bar{1}5)$ reflection in 25-nm-thick single-layer CCO on MAO (Fig.~\ref{fig:XRD}(c)) reveals that the in-plane lattice parameter of the film coincides with that of the substrate, indicating CCO is coherently strained to MAO. 
We note that despite the relatively large lattice mismatch between CCO and MAO of $\approx$3 \%, coherently strained growth of CCO of up to 40 nm has been previously reported on MAO substrates~\cite{Heuver2015}. 
For our CCO film, we calculate an out-of-plane lattice constant $c \approx 8.534$ \AA\ from the $2\theta$-$\omega$ scan; taking the in-plane lattice parameter $a = 8.083$ \AA\ of the MAO substrate, the resulting tetragonal distortion of coherently strained CCO is $c/a \approx 1.055$, similar to that for coherently strained MAFO~\cite{Emori2018a}. 

Structural characterization results underscore the quality of these epitaxial films grown as single layers and bilayers. Considering the comparable high crystalline quality for MAFO, CCO, and MAFO/CCO -- as evidenced by the presence of Laue oscillations and narrow film-peak rocking curves -- we conclude that MAFO/CCO bilayers (with the total thickness limited to $\leq$23 nm) are coherently strained to the substrate. In these samples where the substrate and film layers are isostructural, we also do not expect antiphase boundaries~\cite{Margulies1997, Voogt1998, Hibma1999, Suzuki2001}. Indeed, we find no evidence for frustrated magnetism, i.e., high saturation field and coercivity, that would arise from antiphase boundaries in spinel ferrites~\cite{Margulies1997, Voogt1998, Hibma1999, Suzuki2001}; MAFO/CCO bilayers studied here instead exhibit soft magnetism, i.e., square hysteresis loops with low coercivity $<$0.5 mT, similar to our previous report on epitaxial MAFO thin films~\cite{Emori2018a}.
Thus, MAFO/CCO is a high-quality all-oxide model system, which permits the evaluation of how spin dynamics are impacted by a structurally clean, coherent interface.
%where the crystal lattices of the ferrite and paramagnetic oxide are continuous and isostructural.   

\section{Ferromagnetic Resonance Characterization of Damping}\label{sec:damping}

\begin{figure}[tb]
  \includegraphics [width=1.00\columnwidth] {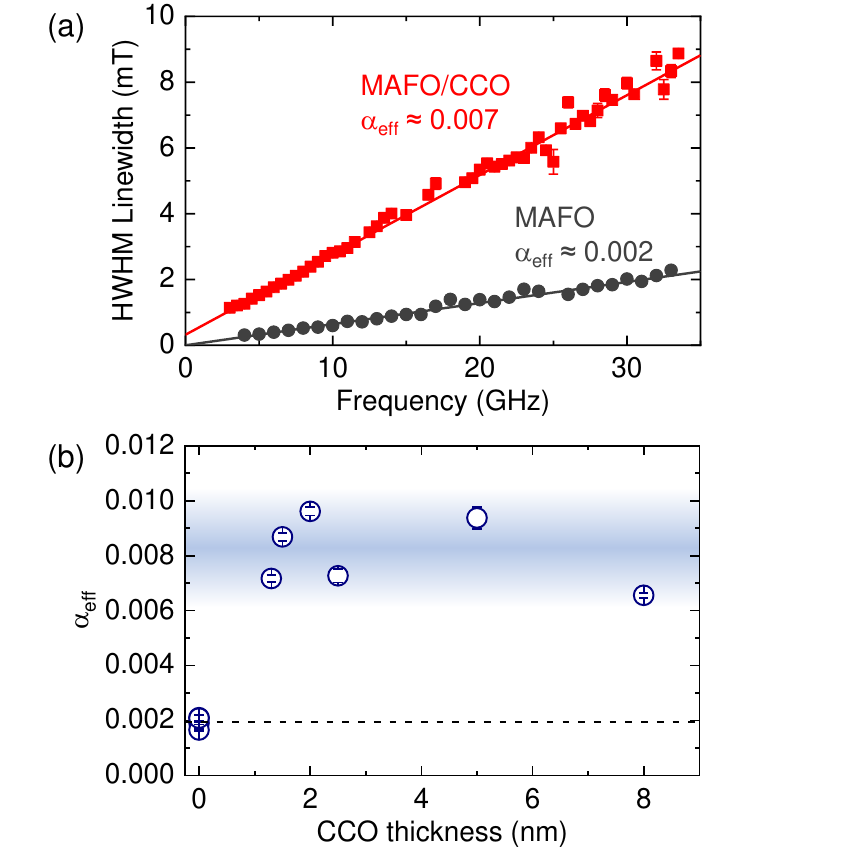}
  \centering
  \caption{\label{fig:damping}
    (a) HWHM FMR linewidth versus frequency for MAFO(15 nm) and MAFO(15 nm)/CCO(1.3 nm). The effective Gilbert damping parameter $\alpha_{eff}$ is derived from the linear fit. (b) $\alpha_{eff}$ plotted against the CCO overlayer thickness. The dashed horizontal line indicates the average of $\alpha_{eff}$ for MAFO without an overlayer.) } 
\end{figure}

To quantify effective damping in coherently strained MAFO(/CCO) thin films, we performed broadband FMR measurements at room temperature in a coplanar waveguide setup using the same procedure as our prior work~\cite{Emori2018a, Emori2017}. We show FMR results with external bias magnetic field applied in the film plane along the [100] direction of MAFO(/CCO); essentially identical damping results were obtained with in-plane field applied along [110]~\cite{Note1}.  
Figure~\ref{fig:damping}(a) shows the frequency $f$ dependence of half-width-at-half-maximum (HWHM) linewidth $\Delta H$ for a single-layer MAFO sample and a MAFO/CCO bilayer with a CCO overlayer thickness of just 1.3 nm, i.e., less than 2 unit cells. The linewidth is related to the effective Gilbert damping parameter $\alpha_{eff}$ via the linear equation:
\begin{equation}
\Delta H=\Delta H_0+\frac{h\alpha_{eff}}{g\mu_0\mu_B}f
\end{equation}
where $\Delta H_0$ is the zero-frequency linewidth, $h$ is Planck's constant, $g \approx 2.05$ is the Land\'e $g$-factor derived from the frequency dependence of resonance field $H_{FMR}$, $\mu_0$ is the permeability of free space, and $\mu_B$ is the Bohr magneton. It is easily seen from Fig.~\ref{fig:damping}(a) that with the addition of ultrathin CCO, the damping parameter is drastically increased, i.e., $>$3 times its value in bare MAFO. 

Figure~\ref{fig:damping}(b) shows that the damping enhancement seen in MAFO/CCO is essentially independent of the CCO thickness. This trend suggests that the damping enhancement is purely due to the MAFO/CCO interface, rather than spin dissipation in the bulk of CCO akin to the absorption of diffusive spin current reported in antiferromagnetic NiO~\cite{Rezende2016,Wang2014g,Ikebuchi2018}. We note that other bulk magnetic properties of MAFO (e.g., effective magnetization, Land\'e $g$-factor, magnetocrystalline anisotropy) are not modified by the CCO overlayer in a detectable way. 
We also rule out effects from solvent cleaning prior to CCO growth or thermal cycling in the deposition chamber up to 300$^\circ$C, as subjecting bare MAFO to the same ex-situ cleaning and in-situ heating/cooling processes as described in Section~\ref{sec:film}, but without CCO deposition, results in no measurable change in damping. The damping enhancement therefore evidently arises from the proximity of MAFO to the CCO overlayer. 

%%%
%%%

We consider two possible mechanisms at the MAFO/CCO interface for the observed damping enhancement: 

(1) Spin current excited by FMR in MAFO may be absorbed via spin transfer in an interfacial proximity-magnetized layer~\cite{Caminale2016} of CCO, whose magnetic moments may not be completely aligned with those of MAFO. While CCO by itself is paramagnetic at room temperature, prior studies have shown that Co$^{2+}$ and Cr$^{3+}$ cations in epitaxial CCO interfaced with a spinel ferrite (e.g., Fe$_3$O$_4$) can develop measurable magnetic order~\cite{Chopdekar2010}. Such damping enhancement due to interfacial magnetic layer is analogous to spin dephasing reported for ferromagnets interfaced directly with proximity-magnetized paramagnetic metal (e.g., Pt, Pd)~\cite{Caminale2016}. 

(2) Even if CCO does not develop proximity-induced magnetism, chemical disorder at the MAFO/CCO interface may enhance spin scattering. For instance, chemical disorder may lead to an increase of Fe$^{2+}$ cations at the MAFO surface, thereby increasing the spin-orbit spin scattering contribution to Gilbert damping in MAFO compared to its intrinsic composition dominated by Fe$^{3+}$ with weak spin-orbit coupling~\cite{Emori2018a,Dionne2011}. Another possibility is that chemical disorder at the MAFO/CCO interface introduces magnetic roughness that gives rise to additional spin scattering, perhaps similar to two-magnon scattering recently reported for ferromagnet/spin-orbit-metal systems~\cite{Zhu2019b}.  
%While two-magnon scattering is usually thought to yield nonlinear frequency dependence of linewidth, ...which is absent here, the large effective magnetization of MAFO may prevent the onset of such nonlinearity in the frequency range measured here. 

In the following section, we directly examine interfacial proximity magnetism and chemical disorder to gain insight into the physical origin of the observed damping enhancement in MAFO/CCO.

\section{Characterization of Interface Chemistry and Magnetism}\label{sec:interface}

To evaluate the potential formation of a magnetized layer in the interfacial CCO through the magnetic proximity effect, we performed depth-resolved and element-specific magnetic characterization of MAFO/CCO bilayers using polarized neutron reflectometry (PNR) and soft magnetic X-ray spectroscopy. PNR measurements were performed using the PBR instrument at the NIST Center for Neutron Research on nominally 15-nm-thick MAFO layers capped with either thick (5 nm) or thin (3 nm) CCO overlayers. PNR measurements were performed in an in-plane applied field of 3 T at temperatures of 300 K and 115 K, the latter case being slightly above the nominal 97 K Curie temperature of CCO~\cite{Heuver2015, Menyuk1962}. Incident neutrons were spin-polarized parallel or anti-parallel to the applied field both before and after scattering from the sample, and the reflected intensity was measured as a function of the perpendicular momentum transfer vector \textbf{Q}. The incident spin state of measured neutrons were retained after scattering, corresponding to the two non-spin-flip reflectivity cross sections ($\uparrow\uparrow$ and $\downarrow\downarrow$). Since all layers of the film are expected to saturate well below the applied field of 3 T, no spin-flip reflectivity is expected and these cross sections were not measured.

\begin{figure}[tb]
  \includegraphics [width=1.00\columnwidth] {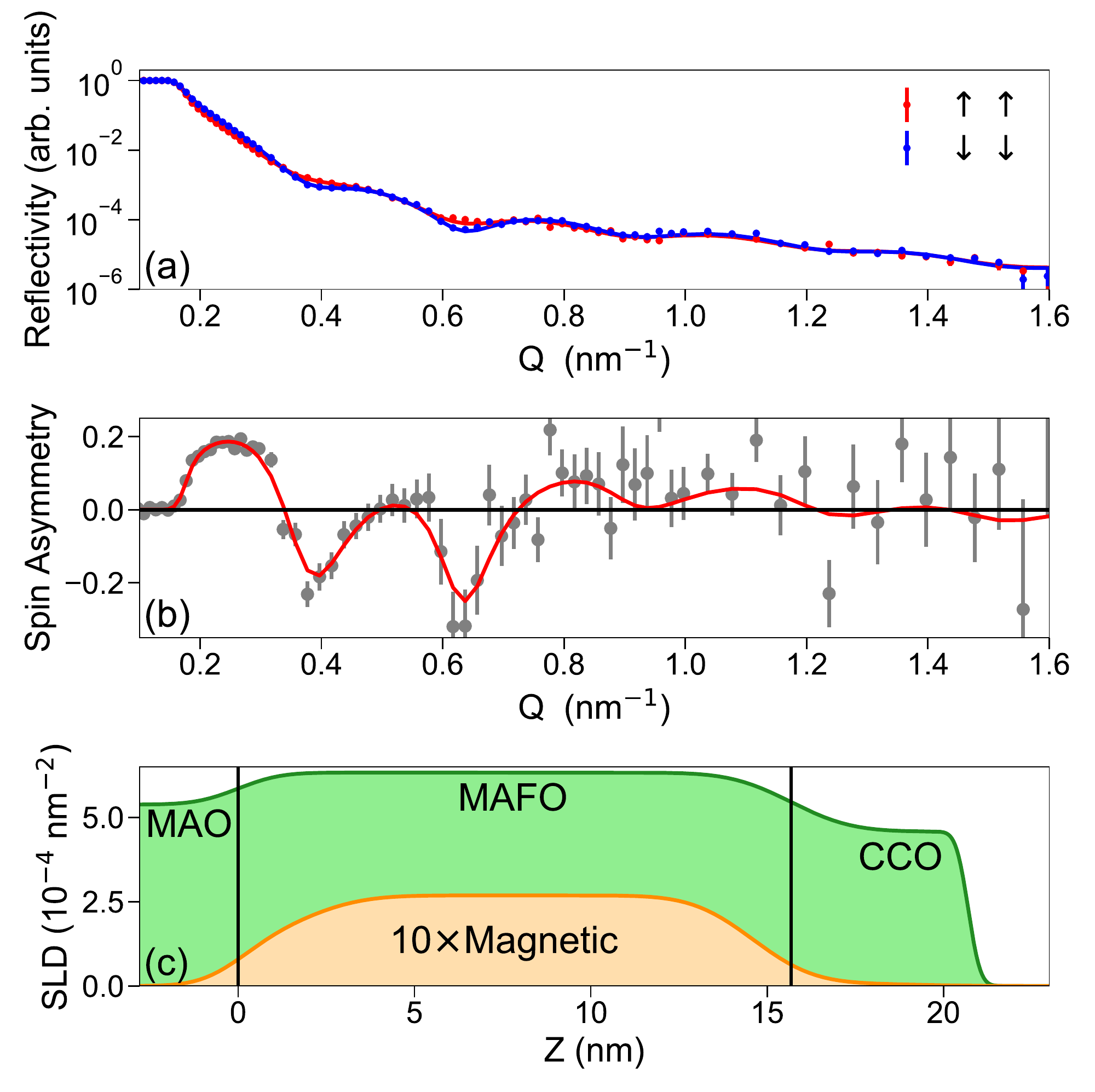}
  \centering
  \caption{\label{fig:PNR}
    (a) Spin-polarized neutron reflectivity and (b) spin asymmetry of a MAFO (15 nm)/CCO (5 nm) bilayer alongside theoretical fits. (c) Nuclear and magnetic scattering (scaled $\times$10) length density profile used to generate the fits shown. Error bars represent $\pm$1 standard deviation.} 
\end{figure}

Since PNR is sensitive to the depth profiles of the nuclear and magnetic scattering length density (SLD), the data can be fitted to extract the chemical and magnetic depth profiles of the heterostructure. In this case, we used the Refl1D software package for this purpose~\cite{Kirby2012}. Figure~\ref{fig:PNR}(a,b) shows the 300 K reflectivities and spin asymmetry curves of a nominal MAFO (15 nm)/CCO (5 nm) sample alongside the depth profile (Fig.~\ref{fig:PNR}(c)) used to generate the fits shown. The best fit profile (Fig.~\ref{fig:PNR}(c)) provides no evidence of a layer with proximity-induced magnetization in the CCO. 
Rather, we note that there appears to be a layer of magnetization suppression near both the MAO/MAFO and MAFO/CCO interfaces. Further, the interfacial roughnesses of both the MAO/MAFO and MAFO/CCO, 0.9(1) nm and 1.35(5) nm respectively, are significantly larger than the CCO surface roughness of 0.27(3) nm and the bare MAFO surface roughness of $\lesssim$0.5 nm~\cite{Wisser2019}. The interfacial roughnesses are signatures of chemical intermixing at the spinel-spinel interface leading to interfacial suppression of the magnetization and/or Curie temperature.
Thus, we find that the MAFO/CCO interface, although structurally coherent, exhibits a chemically intermixed region on the order of one spinel unit cell thick on either side. 

To obtain an upper limit of the proximity-induced interfacial magnetization in CCO, we performed Markov-chain Monte-carlo simulations as implemented in the DREAM algorithm of the BUMPS python package. These simulations suggest an upper limit (95\% confidence interval of) 7 emu/cc in the 1.5 nm of the CCO closest to the interface. In this case, the model evaluated the MAFO as a uniform structural slab but allowed for total or partial magnetization suppression at both interfaces, while the CCO layer was treated as a uniform slab with an allowed magnetization layer of variable thickness at the interface. 

However, we note that equivalently good fits are obtained using simpler models that fit a single MAFO layer with magnetically dead layers at the interfaces and a completely \emph{nonmagnetic} CCO layer. Equivalent results were obtained for the thick CCO sample at 115 K and for the thin CCO sample. We therefore conclude that the PNR results strongly favor a physical picture in which the CCO is \emph{not} magnetized through the magnetic proximity effect.

%%%
%%%

%%%
\begin{figure}[tb]
  \includegraphics [width=1.00\columnwidth] {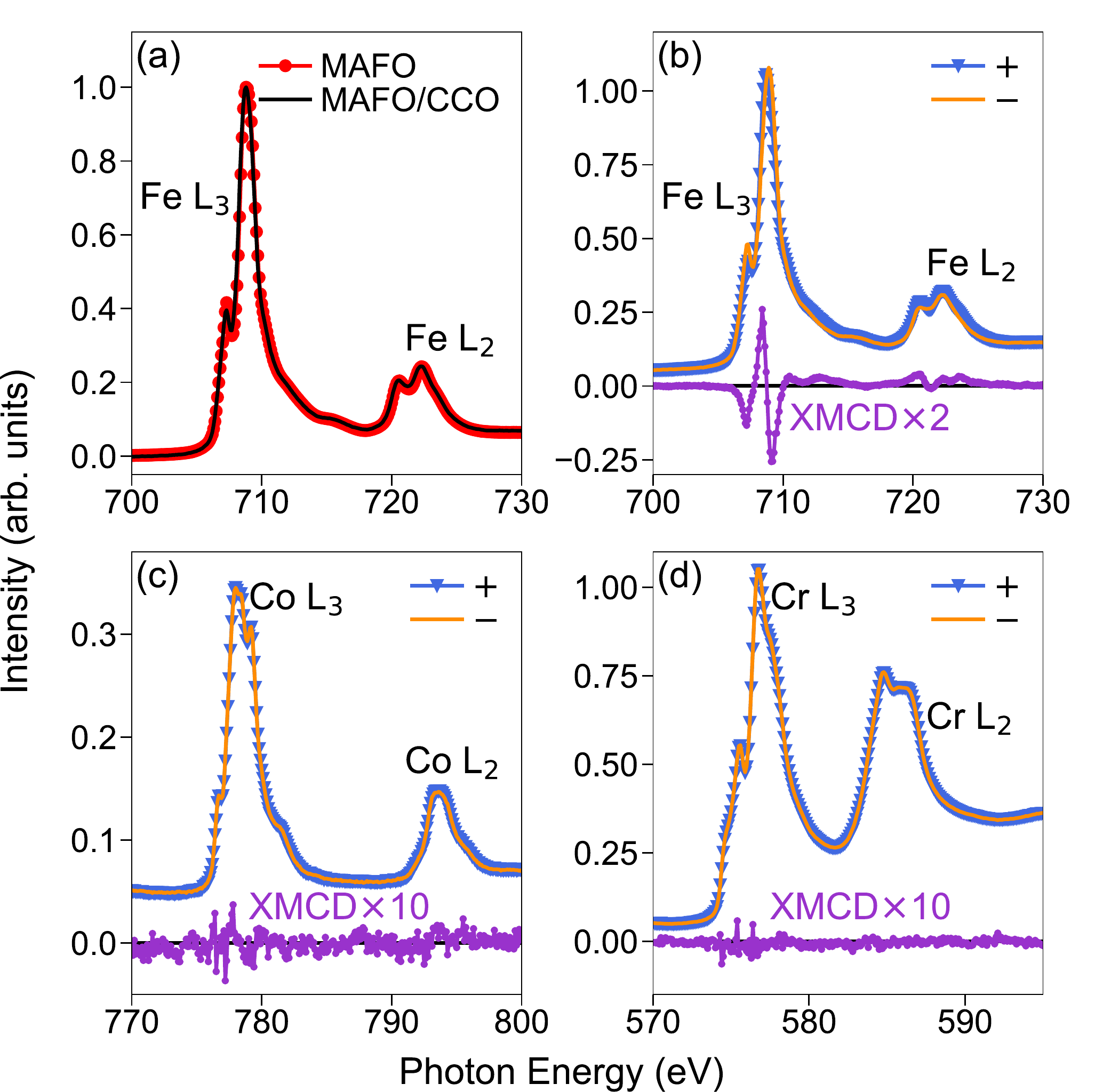}
  \centering
  \caption{\label{fig:xmcd}
    (a) 300 K X-ray absorption spectra of MAFO and MAFO/CCO (3 nm) grown on MAO. (b) Photon helicity-dependent XA spectra and XMCD of the Fe L-edge for a MAFO/CCO (3 nm) bilayer at 300 K. (c) Co and (d) Cr L-edge XA and XMCD of the same bilayer.} 
\end{figure}
%%%

To confirm the PNR results and examine the effect of a CCO overlayer on the local environment of Fe cations in MAFO, we performed temperature-dependent X-ray absorption (XA) spectroscopy and X-ray magnetic circular dichroism (XMCD) measurements at Beamline 4.0.2 of the Advanced Light Source at Lawrence Berkeley National Laboratory. 
We note that the detection mode (total electron yield) used here for XA/XMCD is sensitive to the top $\approx$5 nm of the sample, such that Fe L edge signals from CCO-capped MAFO primarily capture the cation chemistry near the MAFO/CCO interface. 
Measurements were performed in an applied field of 400 mT along the circularly polarized X-ray beam, incident at 30$\degree$ grazing from the film plane.
To minimize drift effects during the measurement, multiple successive energy scans were taken and averaged, switching both applied field direction and photon helicity so that all four possible combinations of field direction and helicity were captured at least once. XA and XMCD intensities were normalized such that the pre-edge is zero and the maximum value of the average of the ($+$) and ($-$) intensities is unity. In the case of the Co L-edge, measurements were taken with energy sweeps covering both Fe and Co edges, and for consistency both edges were normalized to the highest XAS signal, corresponding to the Fe L$_3$-edge. 

Figure~\ref{fig:xmcd}(a) compares the XA of a bare MAFO film with one capped by 3 nm of CCO. The two XA lineshapes are nearly identical, indicating the same average Fe oxidation state and site-distribution in CCO-capped and uncapped MAFO films. It is therefore likely that the reduced interfacial magnetization observed through PNR is a result of a defect-induced Curie temperature reduction, rather than preferential site-occupation of Co and Cr that might increase the Fe$^{2+}$ content in the intermixed interfacial region. 

We further note that although a large XMCD signal is observed on the Fe-edge at 300 K (Fig.~\ref{fig:xmcd}(b)), neither the Co nor Cr L edges exhibit any significant magnetic dichroism, as shown in Figs.~\ref{fig:xmcd}(c)-(d). Similar results are obtained on the Cr L edge at 120 K. Consistent with the PNR results, we thus find no evidence for a net magnetization induced in the CCO through the interfacial magnetic proximity effect.

Our finding of suppressed interfacial magnetism in MAFO/CCO is reminiscent of earlier reports of magnetic dead layers in epitaxially-grown ferrite-based heterostructures~\cite{Ball1996a, Ball1996b,VanderHeijden1996a}. For example, prior PNR experiments have revealed magnetic dead layers at the interfaces of ferrimagnetic spinel Fe$_3$O$_4$ and antiferromagnetic rock-salt NiO or CoO, even when the interfacial roughness is small (e.g., only 0.3 nm)~\cite{Ball1996a, Ball1996b}. A magnetic dead layer of 1 spinel unit cell has also been reported at the interface of Fe$_3$O$_4$ and diamagnetic rock-salt MgO grown by molecular beam epitaxy~\cite{VanderHeijden1996a}. We note that in these prior studies, the spinel ferrite films interfaced with the rock salts (NiO, CoO, MgO) possess antiphase boundaries. Suppressed magnetism is known to result from antiphase boundaries, as they frustrate the long-range magnetic order and reduce the net magnetization of the ferrite~\cite{Voogt1998}. By contrast, there is no evidence for antiphase boundaries in all-spinel MAFO/CCO grown on spinel MAO; therefore, the suppressed magnetism at the MAFO/CCO interface cannot be attributed to antiphase-boundary-induced magnetic frustration. 

Another possible scenario is that magnetic dead layer formation is a fundamental consequence of the charge imbalance between different lattice planes, as recently shown in a recent report of (polar) Fe$_3$O$_4$ undergoing atomic reconstruction to avoid ``polar catastrophe" when grown on (nonpolar) MgO~\cite{Chang2016a}. In our study on all-spinel heterostructures, there may also be some degree of charge mismatch depending on the relative populations of cations on the tetrahedrally- and octahedrally-coordinated sites at the MAFO/CCO interface, although the charge mismatch is expected to be only $\approx$$\pm$1, i.e., a factor of $\approx$5-6 smaller than that in MgO/Fe$_3$O$_4$~\cite{Chang2016a}. Thus, atomic reconstruction driven by charge imbalance appears unlikely as a dominant source of the magnetic dead layer in MAFO/CCO. We instead tentatively attribute the dead layer to atomic intermixing driven by diffusion across the MAFO/CCO interface during CCO overlayer deposition.  

\section{Discussion}
Our PNR and XA/XMCD results (Section~\ref{sec:interface}) indicate that the damping enhancement observed in Section~\ref{sec:damping} arises from chemical disorder, rather than proximity-induced magnetism, at the MAFO/CCO interface. We emphasize that this interfacial disordered layer is confined to within $\approx$2 spinel unit cells. We also note that this interfacial disorder is due to atomic intermixing, but not structural defects (e.g., dislocations, antiphase boundaries), in this coherent bilayer system of MAFO/CCO. Nevertheless, this ultrathin chemically disordered layer alone is evidently sufficient to significantly increase spin scattering. Considering that the cation chemistry of Fe in MAFO does not change substantially (Fig.~\ref{fig:xmcd}(a)), the interfacial spin scattering is likely driven by magnetic roughness, leading to a mechanism similar to two-magnon scattering that accounts for a large fraction of effective damping in metallic ferromagnet/Pt bilayers~\cite{Zhu2019b}. 

We now put in context the magnitude of the damping enhancement $\Delta\alpha_{eff}$, i.e., the difference in the effective Gilbert damping parameter between CCO-capped and bare MAFO, 
\begin{equation}
\Delta\alpha_{eff} = \alpha^{bilayer}_{eff}-\alpha^{ferrite}_{eff}, 
\end{equation}
by comparing it with ferrite/spin-orbit-metal systems where spin pumping is often considered as the source of damping enhancement. 
Since damping enhancement from spin pumping or interfacial scattering scales inversely with the product of the saturation of magnetization $M_s$ and the magnetic layer thickness $t_m$, the values of $\Delta\alpha_{eff}$ taken from the literature~\cite{Sun2013, Wang2014, Riddiford2019} are normalized for direct comparison with the MAFO films studied here with $M_s = 100$ kA/m and $t_m = 15$ nm. 

%%%
\begin{figure}[tb]
  \includegraphics [width=1.00\columnwidth] {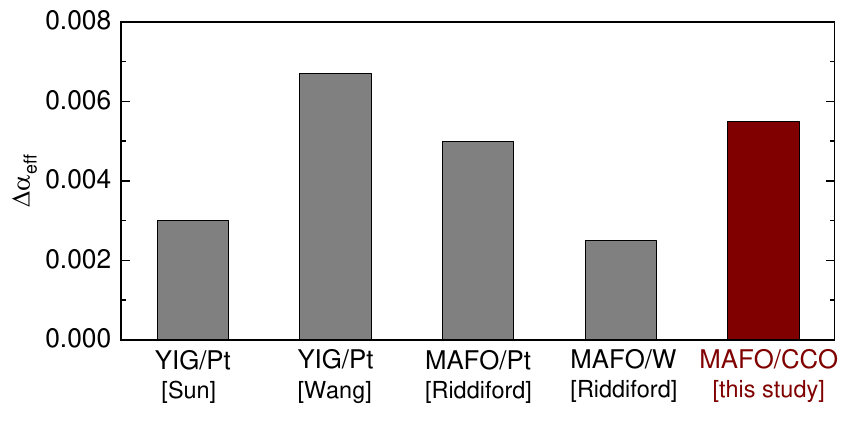}
  \centering
  \caption{\label{fig:lit}
    Comparison of the enhancement of the effective Gilbert damping parameter $\Delta\alpha_{eff}$ for MAFO/CCO and ferrite/spin-orbit-metal bilayers. YIG/Pt [Sun], YIG/Pt [Wang], and MAFO/Pt(W) [Riddiford] are adapted from Refs.~\cite{Sun2013}, \cite{Wang2014}, and \cite{Riddiford2019} respectively. The values of $\Delta\alpha_{eff}$ from the literature are normalized for the saturation magnetization of 100 kA/m and magnetic thickness of 15 nm for direct comparison with our MAFO/CCO result. } 
\end{figure}
%%%

As summarized in Fig.~\ref{fig:lit}, $\Delta\alpha_{eff}$ for MAFO/CCO is comparable to -- or even greater than -- $\Delta\alpha_{eff}$ for ferrite/metal bilayers. This finding highlights that the strength of increased spin scattering in a ferrite due to interfacial chemical disorder can be on par with spin dissipation due to spin pumping in metallic spin sinks. More generally, this finding suggests that special care  may be required in directly relating $\Delta\alpha_{eff}$ to spin pumping across bilayer interfaces (i.e., spin-mixing conductance~\cite{Zhu2019b}), particularly when the FMR-driven magnetic layer is directly interfaced with a spin scatterer. 

Furthermore, the strong interfacial spin scattering -- even when the oxide interface is structurally coherent and the chemically disordered layer is kept to just $\lesssim$2 unit cells -- poses a significant challenge for maintaining low damping in ferrite/insulator heterostructures. 
This challenge is partially analogous to the problem of reduced spin polarization in tunnel junctions consisting of spinel Fe$_3$O$_4$ and oxide barriers (e.g., MgO)~\cite{VanderZaag2000, Alldredge2006, Kado2008, Nagahama2014}, which is also likely due to interfacial chemical disorder and magnetic dead layers. 
However, we emphasize that the problems of antiphase boundaries~\cite{Margulies1997, Voogt1998, Hibma1999, Suzuki2001} and charge-imbalance-driven atomic reconstruction~\cite{Chang2016a}, which have posed intrinsic challenges for devices with MgO/Fe$_3$O$_4$ interfaces, are likely not applicable to all-spinel MAFO/CCO. It is therefore possible that deposition schemes that yield sharper interfaces, e.g., molecular beam epitaxy, can be employed to reduce interfacial imperfections and hence spin scattering at MAFO/CCO for low-loss all-oxide device structures.

\vspace{10pt}
\section{Conclusions}
We have shown that effective damping in epitaxial spinel MgAl-ferrite (MAFO) increases more than threefold when interfaced coherently with an insulating paramagnetic spinel of CoCr$_2$O$_4$ (CCO). This damping enhancement is not due to spin pumping into the bulk of CCO. Our depth-resolved characterization of MAFO/CCO bilayers also reveals no proximity-induced magnetization in CCO or significant change in the cation chemistry of MAFO. We attribute the giant damping enhancement to spin scattering in an ultrathin chemically disordered layer, confined to within 2 spinel unit cells across the MAFO/CCO interface. Our results demonstrate that spin dynamics in ferrite thin films are strongly impacted by interfacial disorder. 

\vspace{10pt}
\textbf{Acknowledgements - } This work was supported in part by the Vannevar Bush Faculty Fellowship program sponsored by the Basic Research Office of the Assistant Secretary of Defense for Research and Engineering and funded by the Office of Naval Research through grant no. N00014-15-1-0045. J.J.W. was supported by the U.S. Department of Energy, Director, Office of Science, Office of Basic Energy Sciences, Division of Materials Sciences and Engineering under Contract No. DESC0008505. Part of this work was performed at the Stanford Nano Shared Facilities (SNSF), supported by the National Science Foundation under award ECCS-1542152.
This research used resources of the Advanced Light Source, which is a DOE Office of Science User Facility under contract no. DE-AC02-05CH11231. We thank Brian J. Kirby for technical assistance on PNR analysis.

\end{document}